\documentclass[fleqn,usenatbib]{mnras}

\usepackage{newtxtext,newtxmath}

\usepackage[T1]{fontenc}

\DeclareRobustCommand{\VAN}[3]{#2}
\let\VANthebibliography\thebibliography
\def\thebibliography{\DeclareRobustCommand{\VAN}[3]{##3}\VANthebibliography}

\usepackage{graphicx}	
\usepackage{amsmath}	
\usepackage{amssymb}	

\usepackage{ulem}
\usepackage{xspace}
\newcommand{\mystar}{WASP-94\,A\xspace}
\newcommand{\myplanet}{WASP-94\,A\,b\xspace}

\newcommand{\locccf}{local CCF\xspace}
\newcommand{\DIccf}{disk-integrated CCF\xspace}
\newcommand{\locccfs}{local CCFs\xspace}
\newcommand{\DIccfs}{disk-integrated CCFs\xspace}
\newcommand{\chisq}{$\chi^{2}$}

\title[HARPS observations of the \mystar system]{
Atmospheric characterisation and tighter constraints on the orbital misalignment of \myplanet with HARPS}

\author[E. Ahrer et al.]{E. Ahrer$^{1,2,3,4}$\thanks{E-mail: ahrer@mpia.de},
J.~V. Seidel$^{5}$\thanks{ESO Fellow, E-mail: jseidel@eso.org},
L. Doyle$^{1,2}$,
S. Gandhi$^{1,2,6}$,
B. Prinoth$^{5,7}$,
H.~M. Cegla$^{1,2}$\thanks{UKRI Future Leaders Fellow}, \newauthor
C. H. McDonald$^{8,1,2}$,
N. Astudillo-Defru$^{9}$, 
E. Ayache$^{10}$, 
R. Nealon$^{1,2}$,
Dimitri Veras$^{1,2,11}$, \newauthor
P. J. Wheatley$^{1,2}$,
D. Ehrenreich$^{12}$
\\
$^{1}$Centre for Exoplanets and Habitability, University of Warwick, Gibbet Hill Road, CV4 7AL Coventry, UK\\
$^{2}$Department of Physics, University of Warwick, Gibbet Hill Road, CV4 7AL Coventry, UK\\
$^{3}$Department of Physics and Astronomy, Faculty of Environment, Science and Economy, University of Exeter, Exeter EX4 4QL, UK \\
$^{4}$Max Planck Institute for Astronomy (MPIA), K\"{o}nigstuhl 17, 69117 Heidelberg, Germany \\
$^{5}$European Southern Observatory, Alonso de C\'ordova 3107, Vitacura, Regi\'on Metropolitana, Chile \\
$^{6}$Leiden Observatory, Leiden University, Postbus 9513, 2300 RA Leiden, The Netherlands\\
$^{7}$Lund Observatory, Division of Astrophysics, Department of Physics, Lund University, Box 43, 221 00 Lund, Sweden\\
$^{8}$Institute of Astronomy, University of Cambridge, Madingley Road, Cambridge, CB3 0HA, UK\\
$^{9}$ Departamento de Matem\'{a}tica y F\'{i}sica Aplicadas, Universidad Cat\'{o}lica de la Sant\'{i}sima Concepci\'{o}n, Alonso de Rivera 2850, Concepci\'{o}n, Chile\\
$^{10}$ The Oskar Klein Centre, Department of Astronomy, Stockholm University, AlbaNova, SE-106 91 Stockholm, Sweden\\
$^{11}$ Centre for Space Domain Awareness, University of Warwick, Gibbet Hill Road, CV4 7AL Coventry, UK \\
$^{12}$Observatoire astronomique de l'Universit\'e de Gen\`eve, Chemin Pegasi 51b, 1290 Versoix, Switzerland\\
}

\date{Accepted XXX. Received YYY; in original form ZZZ}

\pubyear{2024}

\begin{document}
\label{firstpage}
\pagerange{\pageref{firstpage}--\pageref{lastpage}}
\maketitle

\begin{abstract}
We present high spectral resolution observations of the hot Jupiter \myplanet using the HARPS instrument on ESO's 3.6m telescope in La Silla, Chile. We probed for Na absorption in its atmosphere as well as constrained the previously reported misaligned retrograde orbit using the Rossiter-McLaughlin effect. Additionally, we undertook a combined atmospheric retrieval analysis with previously published low-resolution data. 
We confirm the retrograde orbit as well as constrain the orbital misalignment with our measurement of a projected spin-orbit obliquity of $\lambda = 123.0 \pm 3.0 ^\circ$. 
We find a tentative detection of Na absorption in the atmosphere of \myplanet, independent of the treatment of the Rossiter-McLaughlin effect in our analysis (3.6$\sigma$ and 4.4$\sigma$). We combine our HARPS high resolution data with low resolution data from the literature and find that while the posterior distribution of the Na abundance results in a tighter constraint than using a single data set, the detection significance does not improve (3.2$\sigma$), which we attribute to degeneracies between the low and high resolution data.


\end{abstract}




\section{Introduction}

Hot Jupiters  --- Jupiter-sized exoplanets orbiting their host stars at a close distance --- have been at the forefront of exoplanet research as they exhibit relatively large signatures when observing both transit events and obtaining radial velocity measurements due to their closeness to the host star and large size and mass \citep[e.g.][]{Mayor1995AStar,Charbonneau2000DetectionStar}. In addition, they can also show extended atmospheres that result in large atmospheric signatures that can be detected e.g.\ using transmission spectroscopy \citep[e.g.][]{Charbonneau2002DetectionAtmosphere,Snellen2008Ground-based209458b,Snellen2010The209458b,Pont2013TheObservations,Sing2016ADepletion}. Due to their large temperatures hot and ultra-hot Jupiters have dissociated atmospheres, i.e., elements such as Na exist in their atomic and ionic form \citep[e.g.][]{Seager2000TheoreticalTransits}. 

Transmission spectroscopy is the method of determining transit depth versus wavelength and is used to study the atmospheres of exoplanets, identifying molecules and atomic species as well as clouds and hazes that absorb the transmitted light in the observed wavelength ranges. It has been successfully applied in low spectral resolution as well as high spectral resolution regimes. While low resolution spectroscopy allows for studying broadband features e.g.\ scattering slopes and clouds \citep[e.g.][]{LecavelierDesEtangs2008Rayleigh189733b,Kreidberg2014Clouds1214b,Sing2015HSTScattering,Kirk2017RayleighHAT-P-18b, Espinoza2019ACCESS:Magellan/IMACS,Spyratos2021TransmissionWASP-88b,Ahrer2023LRG-BEASTS:B} as well as atomic and molecular signatures such as Na \citep[e.g.][]{Charbonneau2002DetectionAtmosphere,Redfield2008SodiumSpectrumb,Nikolov2018AnExoplanet,Alderson2020LRG-BEASTS:WASP-21b}, water vapour \citep[e.g.][]{Huitson2013AnTiO,Wakeford2015TransmissionExoplanets,Kreidberg2018WaterWASP-107b,Carone2021IndicationsWASP-117b,Ahrer2023EarlyNIRCam,Feinstein2023EarlyNIRISS}, carbon dioxide \citep[e.g.][]{TheJWSTTransitingExoplanetCommunityEarlyReleaseScienceTeam2023IdentificationAtmosphere, Taylor2023} and sulphur dioxide bands \citep[]{Alderson2023EarlyG395H,Rustamkulov2023EarlyPRISM,Dyrek2023so2}, high resolution spectroscopy is used to identify species by resolving their individual lines e.g.\ Fe/Fe+ \citep[e.g.][]{Hoeijmakers2018AtomicKELT-9b,Hoeijmakers2019AB,Ehrenreich2020NightsideExoplanet,Gandhi2022SpatiallyPhase}, Na doublet \citep[e.g.][]{Casasayas-Barris2018KELT-20,Seidel2019HotWASP-76b,Hoeijmakers2020,Prinoth2022WASP-189}, He \citep[e.g.][]{Salz2018HeHD189,Allart2019High-resolutionWASP-107b, Spake2018HeliumExoplanet}, among many others. 

Combining low and high resolution transmission spectroscopy observations allows us to probe for both individual lines as well as continuum in exoplanetary atmospheres \citep[e.g.][]{Brogi2017lowhighres,Pino2018Combining189733b, Khalafinejad2021ProbingSpectroscopy}. For example, \citet{Brogi2017lowhighres} 
showed that combining low resolution data from Hubble Wide Field Camera 3 (WFC3) and high resolution observations with CRyogenic Infra-Red Echelle Spectrograph (CRIRES)  can result in much tighter retrieved constraints on atmospheric abundances in an exoplanet atmosphere, in this case the hot Jupiter HD\,209458\,b. 




In this work, we present high resolution transit observations of \myplanet using the HARPS instrument \citep[High Accuracy Radial Velocity Planet Searcher;][]{Mayor2003SettingHARPS} at ESO's 3.6m-telescope at La Silla, Chile. HARPS has been utilised to detect exoplanets \citep[e.g.][]{Udry2007detection,Bonfils2013harps,Delisle2018,Unger2021TheHD189567} as well as for the characterisation of exoplanet atmospheres \citep[e.g.][]{Wyttenbach2017HotWASP-49b,Seidel2019HotWASP-76b,Mounzer2022,Steiner2023}. 

\myplanet is a hot Jupiter with a mass of half of Jupiter's and a radius of $1.72$\,R$_\mathrm{Jup}$ (see Table\,\ref{tab:wasp-94_parameters}). We further introduce the WASP-94 system in Section\,\ref{sec:wasp-94-system}, followed by a summary of the HARPS observations in Section\,\ref{sec:observations}. We show the Rossiter-McLaughlin effect analysis in Section\,\ref{sec:rm_effect} which includes the updated constraint on the orbital alignment. Our analysis of \myplanet's atmosphere in the form of a transmission spectrum and atmospheric retrieval is presented in Section\,\ref{sec:transmission-spectrum} and \ref{sec:atmospheric-retrieval}. We conclude the paper in Section\,\ref{sec:conclusions}.

\section{WASP-94 system}
\label{sec:wasp-94-system}
The WASP-94 system consists of two F-type stars, \mystar (F8) and WASP-94\,B (F9) with V magnitudes of 10.1 and 10.5, respectively. Their angular distance is $15.03\pm 0.01$~arcseconds and their orbital separation is estimated to be at $2700$\,AU \citep{Neveu-Vanmalle2014WASP-94System}. Using GAIA DR2 the distance of the system was determined to be $212.46 \pm 2.50$~pc \citep[GAIA DR2,][]{Bailer-Jones20182}. 

\citet{Neveu-Vanmalle2014WASP-94System} discovered that both stars host a planetary satellite: \myplanet is transiting and seen in RV measurements while WASP-94\,B\,b does not transit and has been detected by RV measurements only. \myplanet is a hot Jupiter, with a radius of $1.72^{+0.06}_{-0.05}$ R$_\textrm{Jup}$ \citep{Neveu-Vanmalle2014WASP-94System}, a mass of $0.456^{+0.034}_{-0.036}$ M$_\textrm{Jup}$ \citep{Bonomo2017ThePlanets} and an equilibrium temperature of $1508\pm75$\,K \citep{Garhart2020Eclipses}, orbiting its host star in a $3.9501907^{+0.0000044}_{-0.0000030}$ day period \citep{Neveu-Vanmalle2014WASP-94System}. WASP-94\,B\,b is a Jupiter-sized planet with an orbital period of $2.00839 \pm 0.00024$ days \citep{Neveu-Vanmalle2014WASP-94System}. All relevant stellar and planetary parameters of the \mystar system are summarised in Table\,\ref{tab:wasp-94_parameters}. Note that for our modelling of the planetary transit and Rossiter-McLaughlin correction we did not use the planet radius and stellar radius separately as displayed in Table\,\ref{tab:wasp-94_parameters}, instead we used the ratio of planet to star radius $R_\mathrm{p}/R_\mathrm{*} = 0.10859$ in the Na wavelength range from \citet{Ahrer2022LRG-BEASTS:NTT/EFOSC2}.

\begin{table}

    \centering
    \caption{Stellar and planetary parameters for the \mystar planetary system. References are as follows [1] \citet{Neveu-Vanmalle2014WASP-94System},  [2] \citet{Teske2016THEB}, [3] \citet{Bonomo2017ThePlanets}, [4] \citet[GAIA DR3,][]{GaiaDR32023}, [5] \citet[GAIA DR3 GSP-Phot,][]{GaiaDR3GSPPhot} and [6] \citet{Garhart2020Eclipses}. }
    \begin{tabular}{l c c }
    \hline
   \multicolumn{3}{c}{\textbf{ Stellar parameters of \mystar}}\\
    \hline
        Parameter & \mystar &   Reference\\ \hline
        Brightness, $V_{\textrm{mag}}$ & 10.1 & [1]\\
        Spectral type & F8  & [1]\\
        Effective Temperature, $T_{\textrm{eff}}$ (K) & $6194 \pm 5$ & [2]\\
        Age (Gyr) & $2.55 \pm 0.25$  & [2] \\
        Surface gravity, log $g$ (log$_{10}$(cm/s$^{2}$)) & $4.210 \pm 0.011$  & [2]\\
        Metallicity [Fe/H] & $0.320 \pm 0.004$ &  [2] \\
        Mass, $M_\textrm{*}$ ($M_\odot$) & $1.450 \pm 0.090$ & [3], [1]\\
        Radius, $R_\textrm{*}$ ($R_\odot$) & $1.5784^{+0.0095}_{-0.0110}$ & [4,5] \\

        
        Systemic velocity $\gamma$ (km\,s$^{-1}$) & $8.36 \pm 0.19$ & [5] \\

        \hline
   \multicolumn{3}{c}{\textbf{ Planetary parameters of \myplanet}}\\
    \hline
    Parameter & Value  & Reference\\ \hline
        Period, P (days) & $3.9501907^{+0.0000044}_{-0.0000030}$ &[1] \\
        Semi-major axis, a (AU) & $0.0554^{+0.0012}_{-0.0011}$ &[3] \\
        Mass, $M_\textrm{p}$ ($M_{\textrm{Jup}}$) & $0.456^{+0.034}_{-0.036}$ &[3] \\
        Radius, $R_\textrm{p}$ ($R_{\textrm{Jup}}$) & $1.72^{+0.06}_{-0.05}$ &[1] \\
        Inclination, i ($^\circ$) & $88.7 \pm 0.7$ &[1] \\
        Surface gravity, log~g (log$_{10}$(cm/s$^{2}$)) & $2.590^{+0.044}_{-0.042}$ &[3] \\
        Equilibrium temperature, $T_\textrm{eq}$ (K) & $1508 \pm 75$ & [6] 
        \\
    \hline
    \end{tabular}
    
    \label{tab:wasp-94_parameters}
\end{table}

By measuring the Rossiter-McLaughlin (RM) effect \citep{Rossiter1924OnSystem.,McLaughlin1924SomeSystem.} using observations with the CORALIE instrument \citep{Baranne1996ELODIE:Measurements}, \citet{Neveu-Vanmalle2014WASP-94System} found that the orbit of \myplanet is misaligned and likely retrograde with a spin-orbit obliquity of $\lambda = 151^\circ \pm 20^\circ$. 

\myplanet's atmosphere has been previously studied with the EFOSC2 spectrograph on the New Technology Telescope (NTT). \citet{Ahrer2022LRG-BEASTS:NTT/EFOSC2} presented a low resolution transmission spectrum from 4020 -- 7140\,\AA, showing evidence for a scattering slope as well as a Na absorption feature.

\section{Observations}
\label{sec:observations}
Observations of \mystar\ took place as part of the HEARTS survey (ESO programme 097.C-1025(B); PI: Ehrenreich). We collected data of one transit of \mystar b on the night of 15 July 2016 using the HARPS spectrograph \citep{Mayor2003SettingHARPS}. The weather conditions were adequate, slightly affected by thin clouds. The humidity ranged from $30$\,\% at the beginning of the night up to $70\%$ around midnight and down to $40$\,\% at the end of the night. The seeing was not recorded. 

Overall 55 exposures were taken: 5 before the transit, 26 during the transit and 24 after the transit. The airmass values ranged from 2.44 at the beginning of the night, down to 1.004 during the maximum altitude and up to 1.75 at the end of the night. The signal-to-noise (SNR) at the wavelength range of the Na doublet varied throughout the night from 26 up to 62, with a median value of 48. The exposure times for all frames were 600s, except for the first one which was exposed for 700s.

The HARPS Data Reduction Pipeline (DRS v3.8) extracts the spectra order by order from the 2D images from the Echelle spectrograph and are stored as \texttt{e2ds.fits} files. The region of the Na doublet lies in the order number 56, covering a wavelength range of 5850.24 to 5916.17\,\AA.

\section{Rossiter-McLaughlin Analysis}
\label{sec:rm_effect}
When a planet transits a host star, a portion of the starlight is blocked from the line-of-sight and a distortion of the velocities is observed, known as the Rossiter-McLaughlin (RM) effect (see \citet{Rossiter1924OnSystem., McLaughlin1924SomeSystem.} for original studies and \citet{Queloz2000DetectionHD209458} for the first exoplanet case). The shape of the RM curve contains information about the planet-to-stellar radius ratio, the rotational velocity of the star, the impact parameter and the projected obliquity, $\lambda$, which is the sky-projected angle between the stellar spin axis and planetary orbital plane. \citet{Neveu-Vanmalle2014WASP-94System} observed \myplanet using the CORALIE spectrograph \citep{Baranne1996ELODIE:Measurements} obtaining 24 measurements in transit and an additional 24 out-of-transit spectra over the full orbit. Through these observations, they found \myplanet is misaligned and likely retrograde with a projected obliquity of $\lambda = 151 \pm 20^{\circ}$. In this study, we conduct our own analysis of the RM effect using the CORALIE measurements from \citet{Neveu-Vanmalle2014WASP-94System} in combination with our HARPS observations to further constrain the spin-orbit obliquity $\lambda$. 

\begin{figure}
    \centering
    \includegraphics[width = 0.47\textwidth]{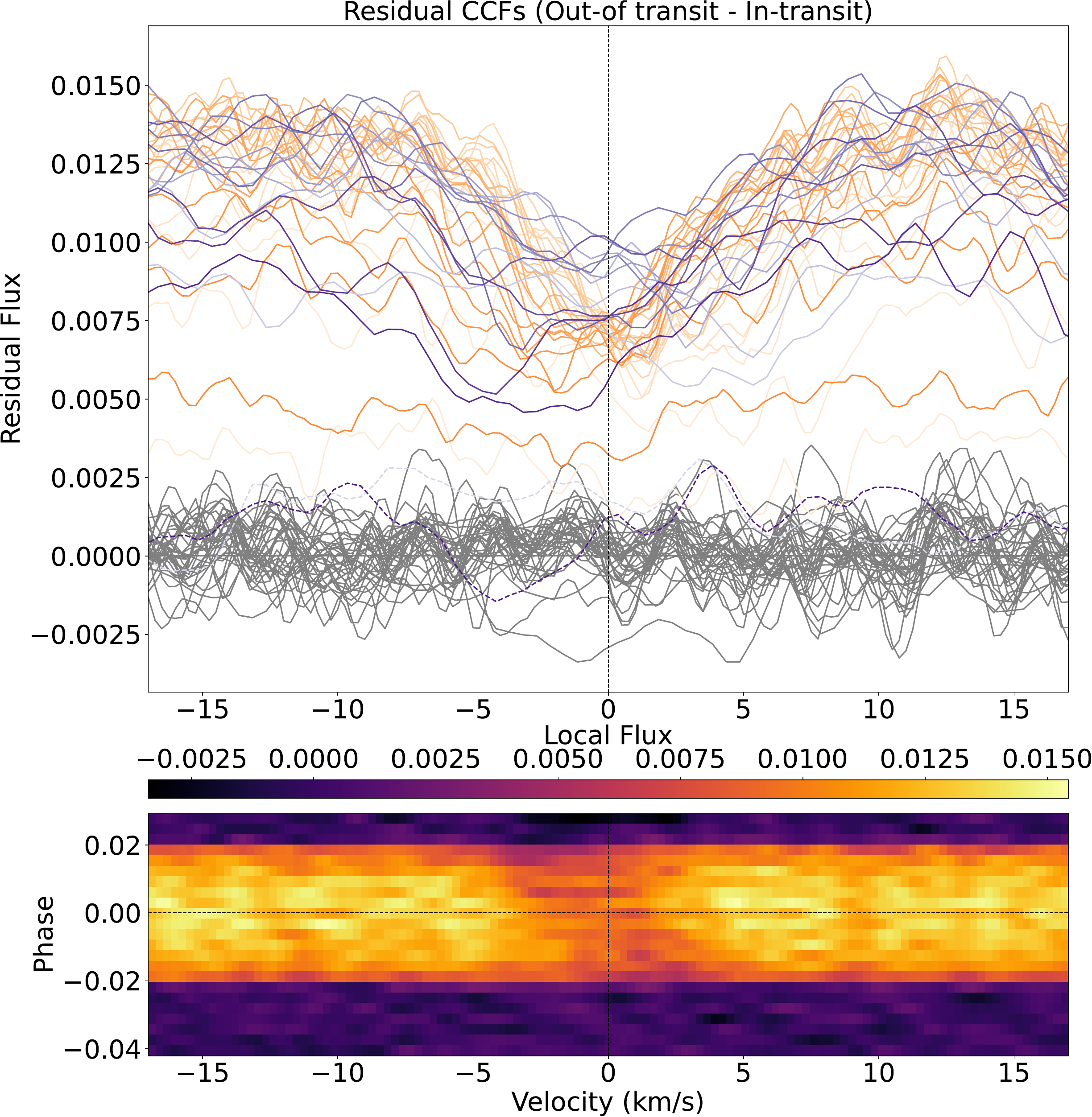}
    \caption{{\it Top panel:} The local CCFs (out-of-transit – in-transit) in the stellar rest frame of the star behind \myplanet. The light grey are the out-of-transit observations and the purple (CORALIE) and orange (HARPS) lines are the in-transit observations. The changing gradient of the orange/purple lines represents the changing centroid position where the darker orange/purple is more redshifted. Dashed orange/purple lines are observations that have a stellar disk position $\langle\mu\rangle <$ 0.30 and are not used in the analysis. 
 {\it Bottom panel:} A top view of all data from both CORALIE and HARPS showing a map of the local CCFs colour-coded by the local flux. A dotted line at phase zero and 0~km\,s$^{-1}$ in both plots is included to guide the eye.   }
    \label{fig:local_ccfs}
\end{figure}

We utilised the Reloaded-RM technique \citep[RRM:][]{Cegla2016TheSystems} which isolates the blocked starlight behind the planet to spatially resolve the stellar spectrum and can be used to derive $\lambda$. Firstly, the CORALIE and HARPS disk-integrated cross-correlation functions (CCFs) are shifted and re-binned in velocity space to correct for the Keplerian motions of the star induced by \myplanet (using the orbital properties in Table\,\ref{tab:wasp-94_parameters}). Next, for each observation a master-out \DIccf was created through a summation of all out-of-transit \DIccfs\ followed by normalising the continuum to unity. Each master-out \DIccf\ was then fitted by a Gaussian profile to determine the systemic velocity, $\gamma$, where this was subtracted from all \DIccfs\ to shift them to the stellar rest frame. Each \DIccf\ was normalised using the continuum and then scaled using a transit light curve model with quadratic limb darkening from the fitted transit parameters in Table\,\ref{tab:wasp-94_parameters}. This allows the \locccfs\ to be obtained by directly subtracting in-transit \DIccfs\ from the master-out \DIccf\ for each night, see Figure\,\ref{fig:local_ccfs}.  

\begin{figure*}
    \centering
    \includegraphics[width = 0.95\textwidth]{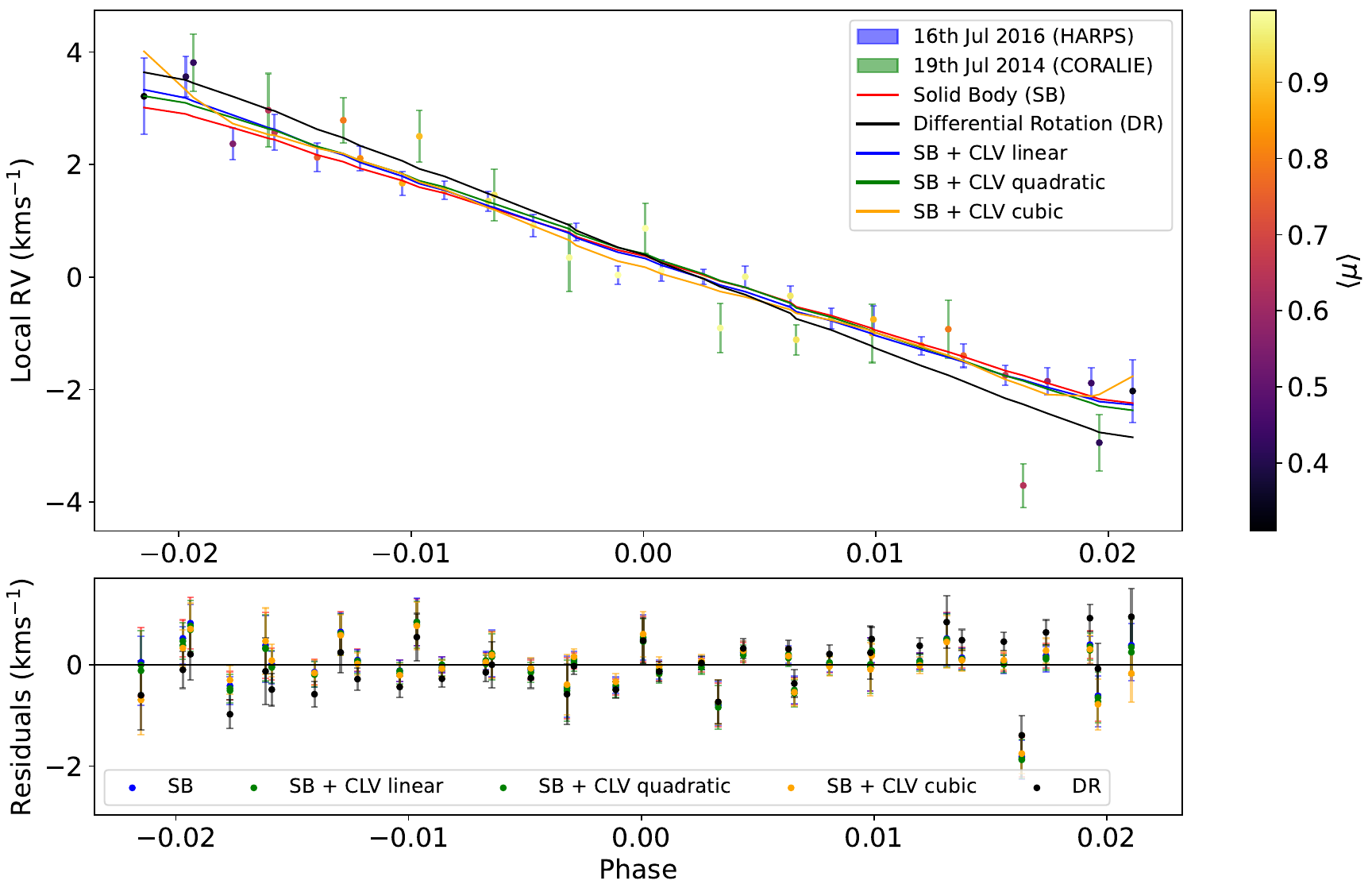}
    \caption{\textit{Top panel:} The local RVs determined from the local CCFs of the regions occulted by the planet as a function of phase. The data points are colour coded by the stellar disk position behind the planet in units of brightness weighted $\langle\mu\rangle$ (where $\mu = \cos\theta$). The best-fit model for solid body rotation (SB: red line) is shown, along with the SB plus centre-to-limb linear (blue), SB plus centre-to-limb quadratic (green), SB plus centre-to-limb cubic (orange) and differential rotation (DR: black) models. \textit{Bottom panel:} The residuals (local RVs - model) for all models with colours corresponding to the top panel model lines, with a horizontal line at 0 to guide the eye.}
    \label{fig:local_rvs}
\end{figure*}

To determine the stellar velocity behind the occulted planet we fit Gaussian profiles to each of the \locccfs\ which included the parameters offset (i.e.\ continuum), amplitude, centroid, and FWHM. Flux errors were determined as the standard deviation of the continuum and were assigned to each \locccf\ and included in the Gaussian fit. The resulting local RVs of the planet occulted starlight can be seen in Figure\,\ref{fig:local_rvs} as a function of phase and stellar disk position behind the planet in units of brightness weighted $\langle\mu\rangle$ (where $\mu = \cos\theta$). We removed CCFs with limb angles $\langle\mu\rangle < 0.30$ from our analysis, resulting in two CCFs being removed from the CORALIE observations. Close to the limb local CCF profiles can be very noisy \citep[see][]{Cegla2016TheSystems}, in this case the depth of the local CCF was not significant enough to enable a Gaussian fit, see Figure\,\ref{fig:local_ccfs} where they are shown as dashed lines. 

The local RVs in Figure\,\ref{fig:local_rvs} were fitted modelling the local, projected stellar velocities behind the planet described in \citet{Cegla2016TheSystems}, henceforth referred to as the RRM model. The model depends on the position of the transiting planet centre with respect to the stellar disk, projected obliquity ($\lambda$), stellar inclination ($i_*$), the equatorial rotational velocity ($v_\mathrm{eq}$), the differential rotational shear ($\alpha$), quadratic stellar limb darkening (u$_1$ and u$_2$), and centre-to-limb convective variations (CLVs) of the star ($v_{\rm{conv}}$). For \mystar, we fitted for different scenarios depending on whether or not we account for differential rotation (DR) and CLVs. We sample the RRM model parameters using a Markov Chain Monte Carlo (MCMC) method implemented using the {\tt Python} package {\tt emcee} \citep{foremanmackey13}. A total of 200 walkers with 5000 steps and an additional burn-in phase of 5000 steps were used. For the solid body (SB) scenario, uniform priors were set on the RRM model parameters $v_\mathrm{eq}\sin i_* \sim \mathcal{U}(0, 100)$ km\,s$^{-1}$ and $\lambda \sim \mathcal{U}(-180^{\circ}, 180^{\circ})$. When including DR, uniform parameters were included for $\alpha \sim \mathcal{U}(-1.0, 1.0)$, $i_* \sim \mathcal{U}(0^{\circ}, 180^{\circ})$, $v_\mathrm{eq} \sim \mathcal{U}(0, 100) $ km\,s$^{-1}$ and $\lambda \sim \mathcal{U}(-180^{\circ}, 180^{\circ})$. Finally, we initiate the walkers in a 
Gaussian ball around the maximum likelihood result. The results are summarised in Table\,\ref{tab:mcmc_results} along with the Bayesian Information Criterion (BIC) for each of the models where a lower value represents a better fit. Note that for all models using SB stellar rotation models $i_*$ and $\alpha$ are fixed under the assumption of rigid body rotation and the $v_{\rm{eq}}$ column corresponds to $v_{\rm{eq}}\sin i_*$. For these models, we are unable to determine the 3D obliquity, $\psi$.

\begin{table*}
    \centering
        \caption{MCMC results for \mystar and the derived 3D spin-orbit obliquity using solid-body (SB) stellar rotation model with centre-to-limb convective variations (CLVs) or differential rotation (DR) model. CLV1, CLV2 and CLV3 correspond to centre-to-limb linear, quadratic and cubic, respectively. The BIC of each model was calculated using \chisq\ and the respective degrees of freedom.} 
    \resizebox{1.0\textwidth}{!}{    
    \begin{tabular}{lcccccccccccc}
    \hline
        Model & No. of Model & $v_{\rm{eq}}$ & $i_{\rm{*}}$ & $\alpha$ & $\lambda$    &  $c_1$        & $c_2$        & $c_3$        & BIC & \chisq &    $\psi$  \\
              & Parameters   & (km\,s$^{-1}$)  & ($^{\circ}$) &          & ($^{\circ}$) &               &              &              &     &        &($^{\circ}$) \\
        \hline
        SB & 2 &  $5.45 \pm 0.45$ & 90.0 & 0.0 & $123.0 \pm 3.0$ & -- & -- & -- & 71.1 & 63.8 &-- \\
        SB + CLV1 & 3 &  $5.94 \pm 0.60$ & 90.0 & 0.0 & $120.2 \pm 3.3$ & $-0.28 \pm 0.25$ & -- & -- & 73.2 & 62.4 &-- \\
        SB + CLV2 & 4 &  $5.95 \pm 0.65$ & 90.0 & 0.0 & $120.6 \pm 3.5$ & $-0.48 \pm 2.0$ & $0.1 \pm 1.3$ & -- & 79.4 & 65.0 &-- \\
        SB + CLV3 & 5 &  $7.07 \pm 0.80$ & 90.0 & 0.0 & $114.8 \pm 3.2$ & $-26.3 \pm 12$ & $38 \pm 17$ & $-17.8 \pm 8.0$ & 75.5 & 57.4 &-- \\
        DR & 4 &  $5.82 \pm \substack{+2 \\ -1}$ & $101.0\substack{+29 \\ -53}$  & $-0.07 \pm 0.25$ & $126.2 \pm 5.0$ & -- & -- & -- & 118 & 103 &-- \\
    
        \hline
    \end{tabular}}
    \label{tab:mcmc_results}
\end{table*}

As \mystar is an F8 type star with an effective temperature of $T_{\rm{eff}}$ = 6194~K, we would expect to observe the net convective velocity shift caused by granules to change as a function of limb angle (i.e.\ from the centre to the limb of the star) due to line-of-sight changes. To account for this, we fit the local RVs for CLV and SB rotation at the same time adding a linear, quadratic or cubic polynomial to the SB model fit as a function of limb angle. However, according to the BIC of the models none of these are significant enough to be considered a detection of CLV. It is highly likely the signal-to-noise of the exposures in both the HARPS and CORALIE data is not high enough to allow for a clear detection \citep[see][for an example with ESPRESSO of a similar spectral type]{doyle2022hot}. Finally, we fit a differential rotation scenario to the local RVs where the result can be seen in Table\,\ref{tab:mcmc_results}. In this case, we find a bimodal distribution to be present in $i_*$ which indicates a degeneracy, likely caused by the spectroscopic transits from HARPS and CORALIE not being precise enough to separate between the star pointing away or towards us. We also ran the MCMC
fitting for DR fixing $i_*$ < 90 degrees (away) and $i_*$ > 90 degrees (towards) to get an estimate on $i_*$ and $\alpha$. Overall, we found these models had a higher BIC than the SB models and $\alpha$ was consistent with zero. Note that we ran additional DR models which include CLV variations and are not listed in Table\,\ref{tab:mcmc_results} as they showed equivalent behaviour to the DR-only model.

In summary, we find the solid body, SB, rotation model is the best fit to the local RVs with a $v_{\rm{eq}}\sin i_*$ = 5.45 $\pm$ 0.45~km\,s$^{-1}$ (from spectral line broadening in \citet{Neveu-Vanmalle2014WASP-94System} $v_{\rm{eq}}\sin i_*$ = 4.2 $\pm$ 0.5~km\,s$^{-1}$, which agrees within 2$\sigma$) and refined projected obliquity of $\lambda$ = 123 $\pm$ 3$^{\circ}$. The projected obliquity lies within 1.5$\sigma$ and the $v_\mathrm{eq}\sin i_*$ lies within 2.5$\sigma$ of the values determined by \cite{Neveu-Vanmalle2014WASP-94System}.

\section{Transmission spectrum}
\label{sec:transmission-spectrum}

We follow \citet{Seidel2019HotWASP-76b} to calculate the transmission spectrum in the order of the Na doublet. We separate the planetary signal from the stellar signal by using the out-of-transit data to create a master spectrum of the star. The in-transit spectra are shifted by the various Doppler shifts moving from the observer's rest frame to the planetary rest frame, in which the results are provided.

\subsection{Telluric correction}

The main impact on transmission spectroscopy from the ground is the absorption of light by Earth's atmosphere which creates additional absorption lines in our spectra. Depending on the relative velocity between the observer and the planet, these lines can overlap with the stellar Na doublet and distort the amplitude and shape of the lines. The established tool to correct for this effect in the optical is \textsc{molecfit} \citep{Smette2015, Kausch2015}, an ESO tool for correcting telluric absorption lines in ground-based high-resolution data (version 1.15.0). We applied the same parameters as used in \cite{Allart2017}. For recent applications of \textsc{molecfit} to the range of the Na doublet for HARPS data see \cite{Hoeijmakers2020, Seidel2020b, Mounzer2022, Steiner2023}. For a more comprehensive overview of different methods to correct tellurics see \cite{Langeveld2021}, with the caveat that they employ an outdated version of \textsc{molecfit} for their assessment.
A visual inspection of the corrected vs. uncorrected spectra showed no over- or under-correction of the telluric lines around the Na doublet. All telluric lines were reduced to the noise level.\\
\noindent A second source of telluric contamination is the emission of telluric Na. The Na layer in Earth's mesosphere can get excited by meteor showers, introducing Na emission features in the Na D doublet. Both the thickness of the layer and the occurrence of meteor showers depend on season and location and thus contamination is not easily predicted. To check for these emission features, fibre B of HARPS is set to the sky and measures the background. We did not find any contamination on fibre B for the presented dataset.


\subsection{Stellar correction}
After the telluric correction was performed, each spectrum was shifted into the stellar rest frame by using the systemic velocity, the barycentric Earth radial velocity (BERV) and the stellar velocity. This was done so that the stellar proportion of each spectrum can be identified as it is constant in the stellar rest frame and eliminated, leaving solely the planet's spectral contribution.

We masked any flux measurements of $<200$ electrons per pixel as data at this level is dominated by read-out noise instead of photon noise. These data thus introduce spurious contamination from the instrument and can potentially mask a planetary signal. A total of 44 pixels were masked, which is an average of 0.8 per exposure. 

While the stellar spectrum does not vary in principle, centre-to-limb variations and the RM effect change our stellar spectra and affect the transmission spectrum. This is likely to be undetectable for slow-rotating stars, but might cause masking of features for stars with rotation \citep{Wyttenbach2020AstronomySeries}. However, different correction methods are currently in use \citep[e.g. ][]{Wyttenbach2020AstronomySeries,Casasayas-Barris2021TheESPRESSO} and we, therefore, test two methods for correcting for these two effects. Note that \mystar\ is also an F-type star where limb variations are not as pronounced as earlier type stars \citep[e.g., see][]{Csizmadia2018limbdarkening}. 

\subsubsection{Numerical correction}
\label{sec:wyttenbach}
First, we followed the stellar correction as described by \citet{Wyttenbach2020AstronomySeries} which uses local stellar spectra (the stellar spectrum behind the planet during the transit) generated from the out-of-transit spectra and the transit depth of the planet to calculate the RM effect and limb-darkening variations for the observed transmission annulus. 

For the transit depth and limb-darkening we utilised the \textsc{batman} \textsc{Python} package \citep{Kreidberg2015BatmanPython} to compute the flux change due to \myplanet\ at each point during the transit. For the local stellar spectra, we used the local RVs computed from RM effect analysis as described in Section\,\ref{sec:rm_effect} to create shifted stellar spectra (from the out-of-transit data) that represent the local stellar spectra at each point during the transit.  

\subsubsection{Modelled correction}
\label{sec:starrotator}
As an alternative and to ensure the accuracy of our RM and CLVs correction, we modelled the effect using the open-source \textsc{StarRotator} \textsc{python} package\footnote{Github: \url{https://github.com/Hoeijmakers/StarRotator}, accessed: 2023-12-07}. Using input star and planetary parameters \textsc{StarRotator} generates the stellar spectrum during the transit. The stellar spectrum was calculated using the Vienna Atomic Line Database \citep[VALD;][]{piskunov1995vald,Ryabchikova2015VALD} in combination with the \textsc{python} installation of \textsc{Spectroscopy Made Easy} \citep[SME;][]{Valenti1996SME,Piskunov2017SMEevolution,Wehrhan2023pySME}, which are commonly used for propagating RM and CLVs effects for high-resolution transmission spectroscopy \citep[e.g.][]{Casasayas-Barris2019Mascara2Kelt20,Yang2023WASP-33}

The star input parameters were based on the preferred solid-body model following our RM analysis (see Table\,\ref{tab:mcmc_results}) and the planet parameters followed Table\,\ref{tab:wasp-94_parameters} except for the transit depth and limb-darkening parameters which were taken from the light curve fit of \myplanet's transit in the Na doublet in low spectral resolution \citep{Ahrer2022LRG-BEASTS:NTT/EFOSC2}. This code then allows us to calculate the RM effect at the spectral resolution of HARPS for the planet phases we observed. We were able to utilise this to correct our in-transit stellar spectra by dividing out the modelled RM effect in each spectrum.

\begin{figure}
    \centering
    \includegraphics[width=\columnwidth]{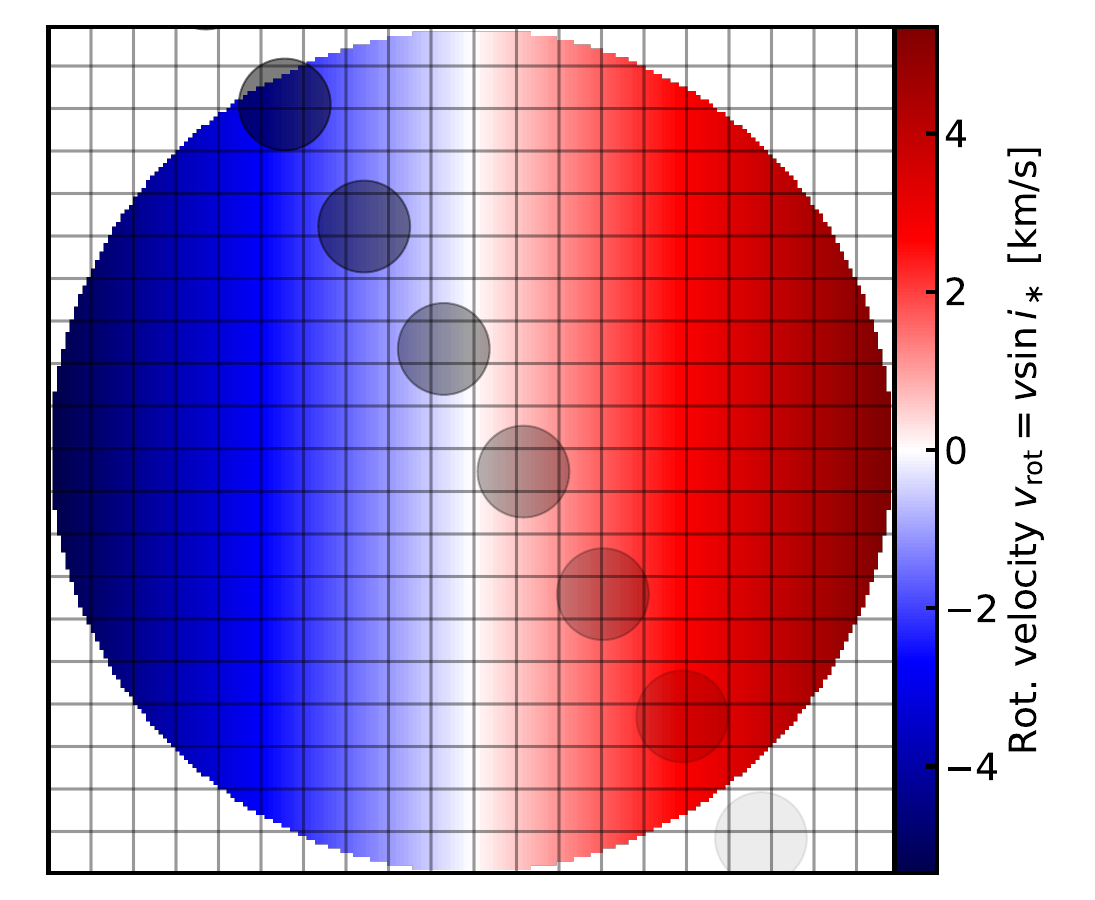}
    \caption{Illustration of the transit of \myplanet, showing the position of the planet at every 4th exposure starting at the bottom right. As it is retrograde it blocks the red-shifted region from the star during the first half of the transit and the blue-shifted area during the second half.  }
    \label{fig:orbital-configuration}
\end{figure}

\subsection{Results}
\subsubsection{Transmission spectrum}
The transmission spectrum was computed by moving each corrected in-transit spectrum into the planet's rest frame, using the planet's velocity at each point in time. These spectra were then summed and normalised by the mean value. We did not take into account any frames where the planetary Na line falls onto the same wavelength bin as the stellar line to avoid any contamination from residuals of the stellar correction. 

\begin{figure}
    \centering
    \includegraphics[width=\columnwidth]{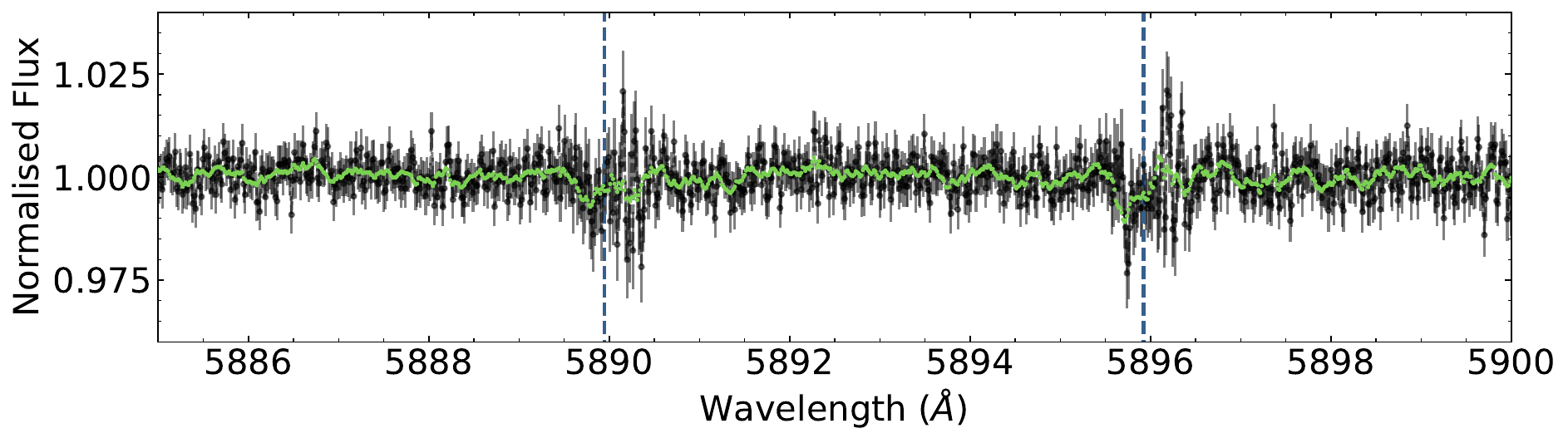}
    \includegraphics[width=\columnwidth]{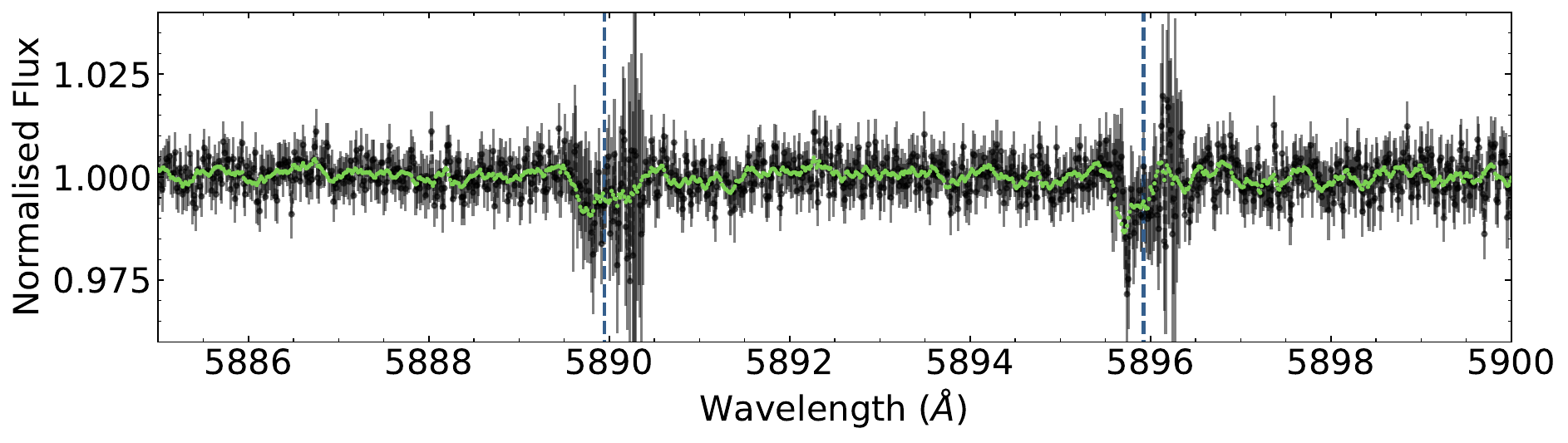}
    \includegraphics[width=\columnwidth]{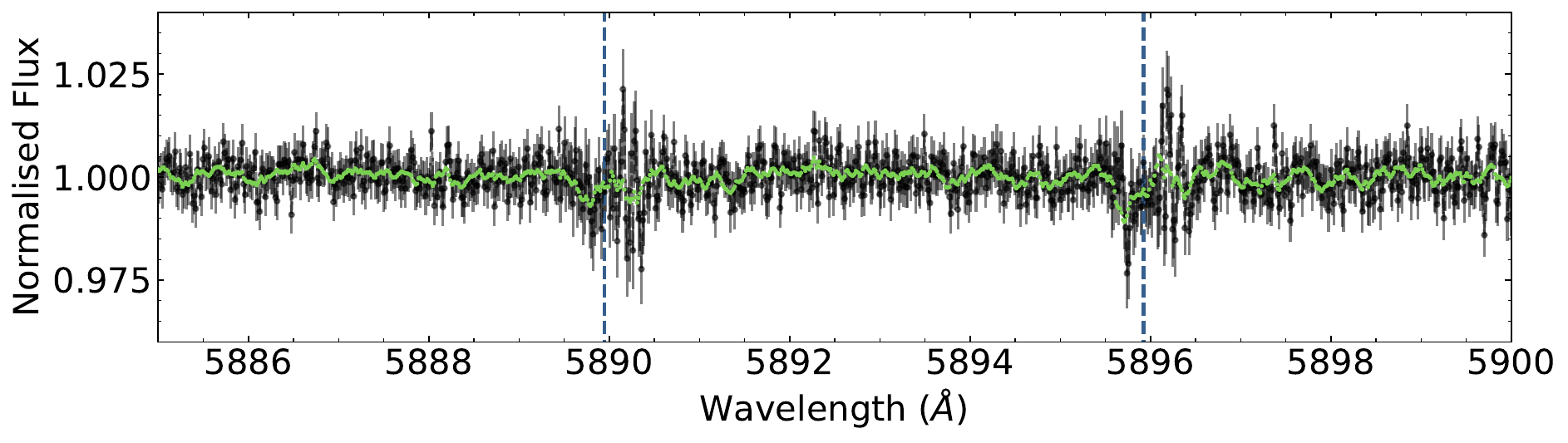}
    \caption{HARPS transmission spectrum of \myplanet in the planetary rest frame, for the wavelength range with the Na doublet. The black points represent the data, while the green line shows the running median over 10 points. The two lines from the Na doublet, D$_2$ and D$_1$, are indicated with the vertical dashed lines at 0.588995 and 0.589592 \textmu m, respectively. \newline \textit{Top panel}: Transmission spectrum without any corrections for the Rossiter-McLaughlin (RM) effect or centre-to-limb variations (CLVs). \newline \textit{Centre panel}: Transmission spectrum using the numerical RM and CLVs correction by \citet{Wyttenbach2020AstronomySeries}. \newline
    \textit{Bottom panel}: Transmission spectrum applying an RM and CLVs correction based on a generated model using \textsc{StarRotator}. Note that the correction is almost non-visible (compared to the top panel), with a maximum effect of 0.5\% at the beginning and the end of the transit which is just about smaller than the average uncertainty.}
    \label{fig:transmission-spectrum}
\end{figure}

We computed the transmission spectrum for three cases, shown in Fig.\,\ref{fig:transmission-spectrum}: (1, top panel) without a correction for the RM effect and centre-to-limb variations, (2, middle panel) RM effect and CLVs is corrected for using the numerical method by \citet{Wyttenbach2020AstronomySeries}, see Section\,\ref{sec:wyttenbach}, and (3, bottom panel) using a \textsc{StarRotator} model to remove effects by the RM and CLVs, see Section\,\ref{sec:starrotator}. 


As a result, in Fig.\,\ref{fig:transmission-spectrum} we see that the Na doublet in the transmission spectrum of \myplanet demonstrates an unusual W-shape using both the numerical and modelled correction for the RM and CLVs. Note that the transmission spectrum using the numerical correction (centre) shows higher uncertainties as we take the uncertainties in the local stellar spectra into account.

\subsubsection{False-positive assessment}
A wrongly identified --- `false-positive' --- detection can occur due to systematic noise e.g.\ from instrumental effects, stellar spots, varying observing conditions, etc. The probability of such a false-positive detection can be estimated using a bootstrapping analysis with an Empirical Monte Carlo (EMC) introduced by \citet{Redfield2008SodiumSpectrumb}. 

For this purpose, we create two independent data sets from our data, where one includes the in-transit data while the other consists of the out-of-transit data, all in the planetary rest frame. Following \citet{Redfield2008SodiumSpectrumb}, we randomly draw in-transit and out-of-transit data sets to create a virtual transmission spectrum. We consider three scenarios: (1) `in-in', where all drawn spectra are taken from the real in-transit data and randomly considered either in- or out-of-transit for our virtual transmission spectrum; (2) `out-out', where all spectra are drawn from the real out-of-transit data and randomly considered as either in- or out-of-transit for our virtual transmission spectrum; (3) `in-out', where the virtual transmission spectrum is created using randomly drawn real in-transit and out-of-transit spectra for the virtual in- and out-of-transit spectra. In the case where the detection is not spurious and indeed of planetary nature we expect that only the `in-out' scenario shows a detection. 

\begin{figure*}
    \centering
    \begin{minipage}{0.47\textwidth}
    \includegraphics[width=\textwidth]{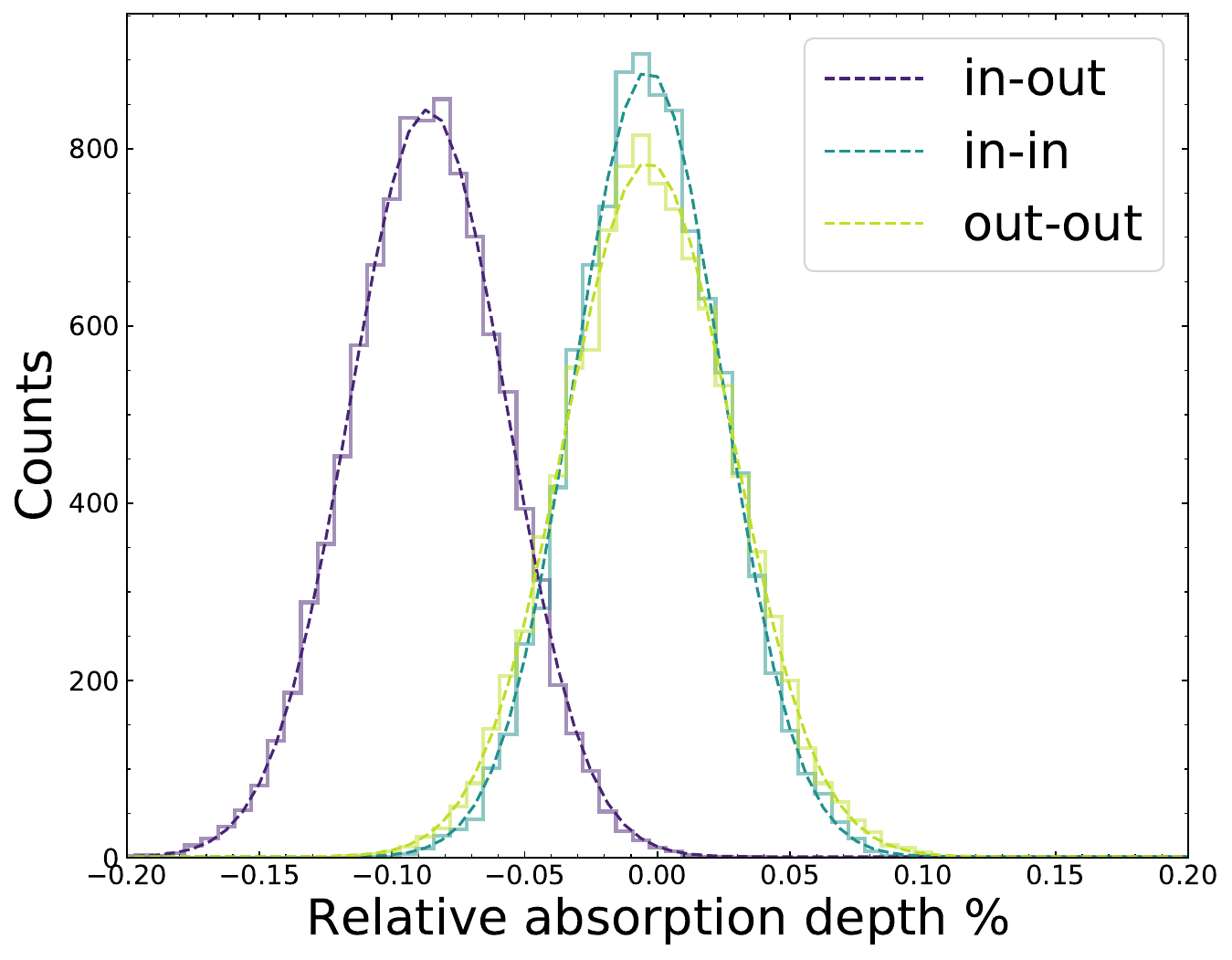}
    \end{minipage}
    \hfill
    \begin{minipage}{0.47\textwidth}
    \includegraphics[width=\textwidth]{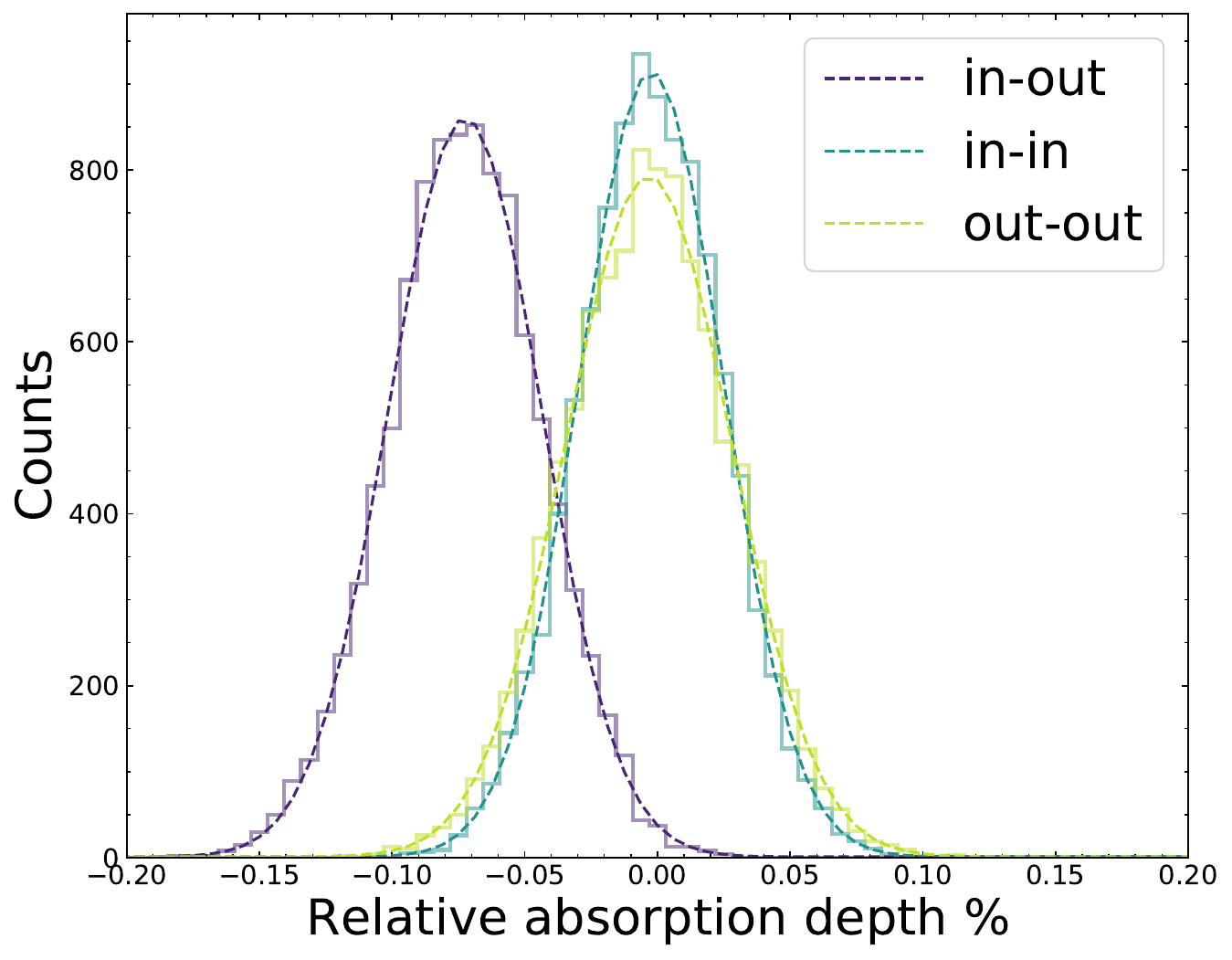}
    \end{minipage}
    \caption{Distribution of our bootstrapping analysis for the Na doublet for 5,000 random selections for the stellar-corrected spectra using the numerical RM correction (left) and using the \textsc{StarRotator} modelled correction (right). The `in-out' describes the case where an in- and out-of-transit spectrum were used, while the `in-in' and `out-out' describe the cases where only in- and only out-of-transit spectra were utilised, respectively. Therefore the `in-out' distribution is the only one that is expected to be centred at a non-zero value if the signal is indeed of planetary origin which is true in both cases, slightly stronger using the numerical RM correction (left).}
    \label{fig:bootstrapping}
\end{figure*}

The results of this bootstrapping assessment are shown in Fig.\,\ref{fig:bootstrapping}, where we conducted this analysis using the in-transit spectra when correcting using the numerical correction (Section\,\ref{sec:wyttenbach}) and the \textsc{StarRotator} modelled correction (Section\,\ref{sec:starrotator}). Both of them show that `in-in' and `out-out' are centred at zero, while the `in-out' is non-zero with a fitted peak at $\approx-0.1 \%$ and a standard deviation of $\approx 0.03 \%$, thus ruling out a spurious detection of a planetary atmosphere due to systematic effects.


\subsubsection{Na absorption depth}
\label{sec:sodium-depth}

\begin{figure*}
    \centering
    \includegraphics[width=\textwidth]{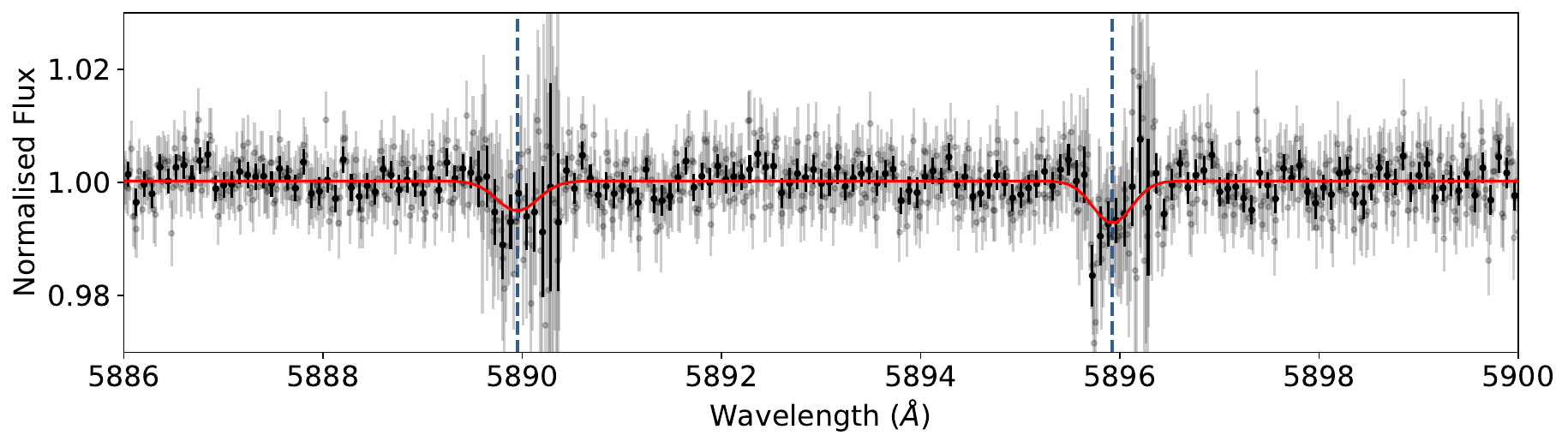}
    \includegraphics[width=\textwidth]{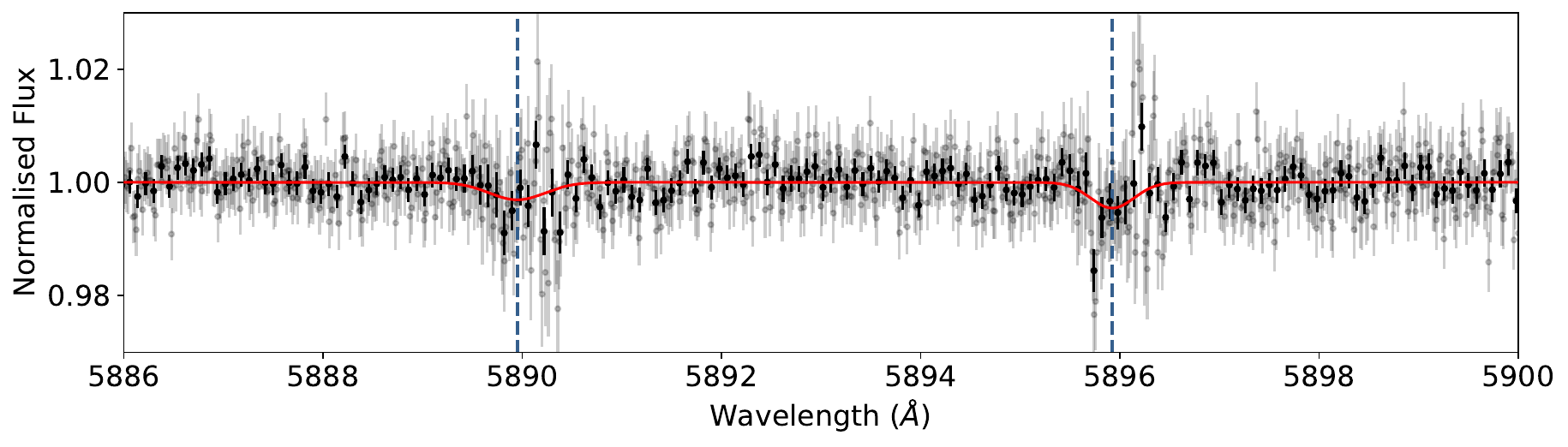}
    \caption{\myplanet\ transmission spectrum binned in black and unbinned in light grey with a Gaussian fit to both Na doublet lines in red. \textit{Top panel:} Transmission spectrum retrieved with numerical stellar correction method (Section\,\ref{sec:wyttenbach}) and its fitted Gaussian function. \textit{Bottom panel:} Transmission spectrum retrieved with modelled stellar correction (Section\,\ref{sec:starrotator}) and its Gaussian fit.  }
    \label{fig:spectrum-fit}
\end{figure*}

We fitted Gaussian profiles to both lines in the Na doublet of the two transmission spectra retrieved with the numerical and modelled stellar correction using a nested sampling algorithm \citep[\textsc{Polychord}][]{Handley2015PolyChord:Sampling}. The resulting fits are shown in Fig.\,\ref{fig:spectrum-fit}. The Gaussian fit to both lines is strongly favoured over a straight line with a Bayesian evidence difference of $9.5$ ($\approx 4.0\sigma$) and $8.5$ ($\approx 2.7 \sigma$) for the numerical and modelled approach, respectively. 


We calculated the relative absorption depth by averaging the flux in the Na lines and comparing it to the average flux in chosen reference bands in the continuum, following e.g.\ \citet{Charbonneau2002DetectionAtmosphere}. We selected a blue (B) and red (R) control wavelength band with a width of 12 \AA, at 5874 --- 5886 \AA\ and 5898 --- 5910 \AA, respectively. The fluxes at the Na doublet are then summed within an area of 1.5\,\AA, 3\,\AA, 6\,\AA\ and 12\,\AA\ on each side of each line core \citep[following e.g.][]{Wyttenbach2015SpectrallySpectrograph,Wyttenbach2017HotWASP-49b,Seidel2019HotWASP-76b} and subtracted from the average of the fluxes in the B and R wavelength bands. The absorption depths found are displayed in Table\,\ref{tab:absorption-depths} with their respective detection levels. For the two narrowest Na bands the depths range from $0.118 \%$ up to $0.209 \%$ with detection significances of $2.7 \sigma$ up to $4.4 \sigma$. This is also consistent with the significance found using the Bayesian evidence model differences between fitting a straight line and fitting Gaussian functions to the absorption lines discussed at the beginning of this Section.

\begin{table}
    \centering
    \caption{Relative absorption depth and detection level of Na in the atmosphere of \myplanet (calculated as depth over the standard deviation), observed with HARPS for different wavelength band widths.}
    \label{tab:absorption-depths}
    \begin{tabular}{ccccc}
    \hline
         & \multicolumn{2}{c}{\textbf{Numerical correction}}  & \multicolumn{2}{c}{\textbf{Modelled correction}}\\ \hline
         band & depth ($\%)$ & $\sigma$ & depth ($\%)$ & $\sigma$\\ \hline
         2 $\times$ 0.75 \AA & $0.209 \pm 0.058$ & $3.7\sigma$ &  $0.149 \pm 0.055$ & $2.7\sigma$ \\
         2 $\times$ 1.5 \AA & $0.147 \pm 0.034$ & $4.4\sigma$ &  $0.118 \pm 0.033$ & $3.6\sigma$ \\
         2 $\times$ 3 \AA & $0.062 \pm 0.023$ & $2.8\sigma$ &  $0.047 \pm 0.022$ & $2.2\sigma$ \\
         2 $\times$ 6 \AA & $0.040 \pm 0.018$ & $2.3\sigma$ &  $0.028 \pm 0.017$ & $1.7\sigma$ \\ \hline
    \end{tabular}
\end{table}

\section{Atmospheric Retrieval}
\label{sec:atmospheric-retrieval}

We perform atmospheric retrievals with HyDRA-H \citep{Gandhi2019HyDRA-H:Spectra, Gandhi2022SpatiallyPhase}, combining the high spectral resolution observations with HARPS using the numerical stellar correction (as it showed the larger Na feature and is based on our real data) with low resolution observations taken with the NTT \citep{Ahrer2022LRG-BEASTS:NTT/EFOSC2}. Our setup includes three free parameters for the volume mixing ratios of H$_2$O, Na and K, with their opacities derived using the ExoMol line list for H$_2$O \citep{Polyansky2018ExoMolWater}, and the Kurucz line list for Na and K \citep{Kurucz1995AtomicData}. We additionally include 6 parameters that determine the temperature profile of the atmosphere using the method of \citet{Madhusudhan2009AAtmospheres}. We also include a reference pressure at which the planetary radius is set into our retrieval. We allow for partial clouds and hazes in the atmosphere of \myplanet, using the cloud fraction as a free parameter \citep{Line2016THESPECTRA}. For the high resolution observations, we also include two additional parameters, the deviation from the planet's known systemic velocity, dV$_\mathrm{sys}$, and a term to account for additional broadening introduced by the rotation velocity of the planet. Overall, our retrieval has 16 free parameters, with the prior ranges shown in Table~\ref{tab:priors}. 

\begin{table}
    \centering
    \caption{Parameters and uniform prior ranges for our retrieval. We retrieve the Na, K and H$_2$O abundances, temperature profile, and partial cloud/haze parameters. Our temperature profile includes 6 free parameters, and our cloud/haze parametrisation includes 4 free parameters (see Section\,\ref{sec:atmospheric-retrieval}). The quoted values retrieved values here are from using both high resolutions HARPS and low resolution NTT data.}
    \label{tab:priors}
    \def\arraystretch{1.5}
    \begin{tabular}{c|c|c}
    \hline
\textbf{Parameter}              & \textbf{Prior Range} & \textbf{Retrieval Constraint}\\
\hline
$\log(X_\mathrm{H_2O})$  & $\mathcal{U}(-15, -1)$& $-8.4^{+4.1}_{-4.0}$\\
$\log(X_\mathrm{Na})$ & $\mathcal{U}(-15, -1)$ & $-6.6^{+1.6}_{-1.0}$\\
$\log(X_\mathrm{K})$ & $\mathcal{U}(-15, -1)$ & $-6.7^{+2.0}_{-4.9}$\\
$T_\mathrm{top}$ / K & $\mathcal{U}(750, 3000)$ & $1900^{+680}_{-1270}$\\
$\alpha_1\, /\, \mathrm{K}^{-\frac{1}{2}}$ & $\mathcal{U}(0, 1)$& $0.7 \pm 0.2 $\\
$\alpha_2\, /\, \mathrm{K}^{-\frac{1}{2}}$ & $\mathcal{U}(0, 1)$ & $0.6^{+0.3}_{-0.2}$\\
$\log(P_1 / \mathrm{bar})$ & $\mathcal{U}(-6, 2)$ & $-1.6 \pm 1.6$\\
$\log(P_2 / \mathrm{bar})$ & $\mathcal{U}(-6, 2)$ & $-4.1^{+1.6}_{-1.2}$\\
$\log(P_3 / \mathrm{bar})$ & $\mathcal{U}(-2, 2)$ & $0.6^{+0.9}_{-1.2}$\\
$\log(P_\mathrm{ref} / \mathrm{bar})$ &$\mathcal{U}(-4, 2)$ & $-3.4^{+0.7}_{-0.9}$\\
$\log(\alpha_\mathrm{haze})$ & $\mathcal{U}(-4, 6)$ & $1.7^{+2.1}_{-3.3}$\\
$\gamma_\mathrm{haze}$ & $\mathcal{U}(-20, -1)$ & $-12.3^{+6.3}_{-5.0}$\\
$\log(P_\mathrm{cl}/\mathrm{bar})$ & $\mathcal{U}(-6, 2)$ & $-3.2^{+1.4}_{-1.2}$\\
$\phi_\mathrm{cl}$ & $\mathcal{U}(0, 1)$  & $0.7^{+0.2}_{-0.3}$\\ 
dV$_\mathrm{sys}$ & $\mathcal{U}(-50, 50)$ & $-2.1^{+6.2}_{-5.4}$\\
Rotation rate/ km\,s$^{-1}$ & $\mathcal{U}(0, 50)$  & $12.2^{+11.2}_{-5.7}$\\
\hline
    \end{tabular}
    
\end{table}

\begin{figure}
    \centering
    \includegraphics[width=\columnwidth]{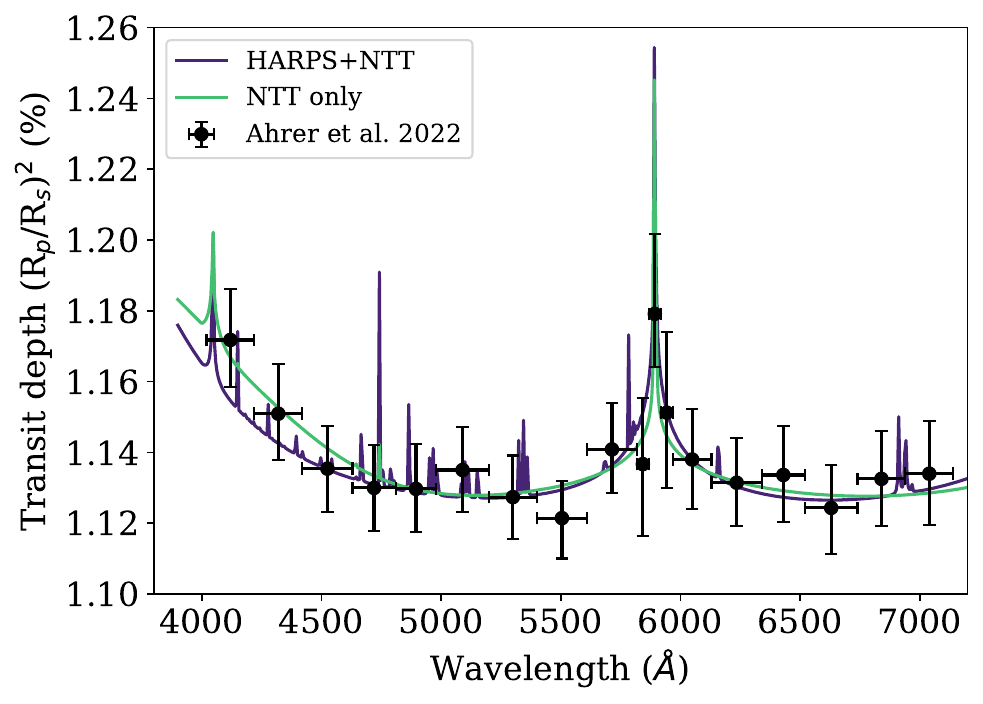}
    \caption{Low spectral resolution transmission spectrum of \myplanet\ using NTT/EFOSC2 (black) as published by \citet{Ahrer2022LRG-BEASTS:NTT/EFOSC2} with our retrieved models using NTT data only (NTT only, turquoise) and combining it with our HARPS observations (HARPS+NTT, dark blue).  }
    \label{fig:low-res}
\end{figure}

We perform three retrievals, one using both data sets as well as one for each data set individually. The retrieved parameters using both low and high resolution data are shown next to their priors in Table\,\ref{tab:priors}. The low resolution NTT/EFOSC2 data is shown in Fig.\,\ref{fig:low-res} along with the combined and individual retrieved model. 

\begin{figure}
    \centering
    \includegraphics[width=\columnwidth]{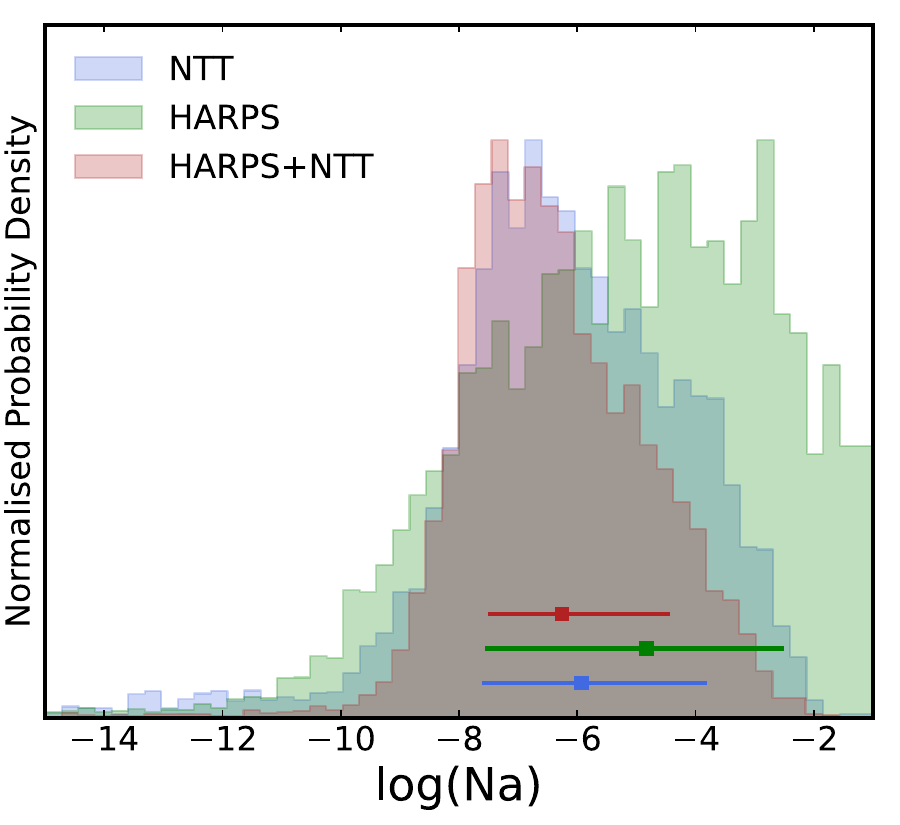}
    \includegraphics[width=\columnwidth]{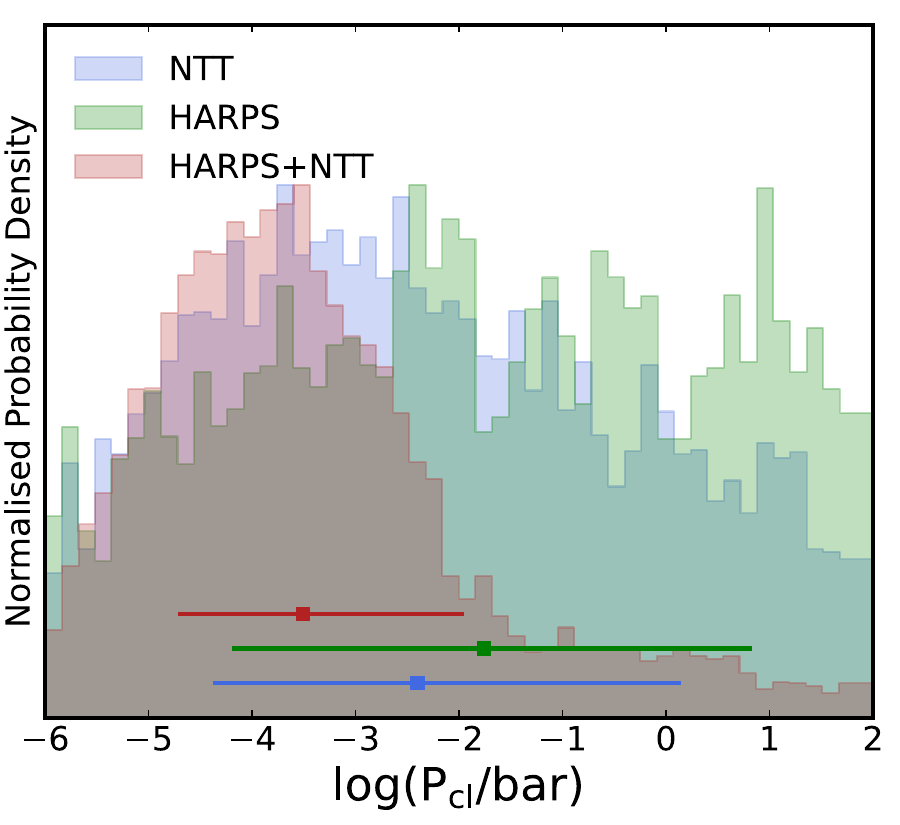}
    \caption{Posterior distributions of the atmospherical retrieval analysis using the HARPS observations from this work, applying a numerical correction to the RM-effect which showed higher detection significance than the modelled approach (see Section\,\ref{sec:sodium-depth} and Table\,\ref{tab:absorption-depths}) and combined with the low spectral resolution data (NTT) from \citet{Ahrer2022LRG-BEASTS:NTT/EFOSC2}. \textit{Top panel:} Constraints on the abundance of Na from NTT, HARPS and combined. \textit{Bottom panel:} Constraints on the cloud-top level from NTT, HARPS and combined. }
    \label{fig:retrieval-posteriors}
\end{figure}

While most parameters were consistent across all retrievals, the abundance of Na and cloud-top level show tighter constraints when combining the low and high spectral resolution from NTT and HARPS, as seen in Fig.\,\ref{fig:retrieval-posteriors}.
However, when computing the detection significance for Na by comparing the Bayesian evidence for models with and without Na, we did not find a significant improvement when retrieving Na on both low and high resolution data sets with 2.6$\sigma$ (NTT only), 3.1$\sigma$ (HARPS only) and 3.2$\sigma$ (NTT+HARPS combined).

\section{Discussion and Conclusions}
\label{sec:conclusions}
\subsection{Orbital misalignment}
With our analysis of the Rossiter-McLaughlin effect, we have confirmed the previously suggested retrograde orbit of \myplanet\ \citep{Neveu-Vanmalle2014WASP-94System} and retrieved constraints on the projected spin-orbit obliquity of $\lambda = 123 \pm 3 ^\circ$. The orbital misalignment of \myplanet allows us to draw some conclusions about the system's formation and evolution.

Its close-in orbit makes it unlikely that \myplanet formed around the stellar companion, WASP-94\,B, and was dynamically exchanged between the two stars \citep{MoeckelVeras2012:BinaryOrbitHop}, or that its highly inclined orbit could be explained by capture of a free-floating planet in analogy to the retrograde orbit of Neptune's moon Triton \citep{AgnorHamilton2006:TritonCapture, GoulinksiRibak2018:PlanetCapture}.

Therefore, \myplanet likely formed in situ around \mystar and then migrated to its current orbital configuration.
Models in which planets migrate within the protoplanetary disc achieve lower obliquities than that required here \citep{LinPapaloizou1986:PPDTides,Millholland2020tidalrunaway}, suggesting that \myplanet migrated after its protoplanetary disc had dissipated. 
Similarly, perturbations driven by planetary mass companions also struggle to achieve high obliquities \citep{Chatterjee2008:planetplanetscatter, PetrovichTremaine2016:WarmJupiters}.

The stellar companion WASP-94\,B is then likely the origin of the perturbation which drove \myplanet towards its current orbit (see figure 24 of \citealt{Albrecht2022:StellarObliquities} for a comparison of the obliquity distributions produced by various misalignment mechanisms). 
The well-studied Kozai-Lidov mechanism \citep{Kozai1962, Lidov1962} involves a distant companion driving alternating cycles of high eccentricity and inclination for an inner companion and has been widely invoked to form hot Jupiters \citep[e.g.][]{Holman1997:16CygniB, WuMurray2003:HD80606, Naoz2011:HJs}.
Kozai-Lidov cycles driven by a stellar companion can force planets onto high obliquity and even retrograde orbits \citep{Naoz2011:HJs, Anderson2016:HJ_StellarKozai, Li2014:KL_OrbitFlip}.
The planet's pericentre can be driven to small enough distances that tidal interactions between the planet and star can disrupt the cycle and cause the planet's orbit to shrink and circularize at its frozen in orientation \citep[e.g.][]{RasioFord1996:SystemFormation, FabryckyTremaine2007:KozaiTidal, Matsumura2010:TidalEvolution}, producing a misaligned hot Jupiter.

The effective temperature of \mystar \citep[$T_\text{eff} = 6194 \pm 5$\,K, ][]{Teske2016THEB} places it close to the Kraft break, which is a transition between fast rotating hot stars and cooler stars with thicker convective envelopes and slower rotations due to magnetic braking \citep{Kraft1967:RotationBreak, Albrecht2022:StellarObliquities, Dawson2014:HJTidal}.
Previous studies have identified that stars above $T_\text{eff} = 6000$~K host hot Jupiters with a wider range of obliquities than cooler stars \citep{Winn2010:HotObliquities, Schlaufman2010:HotObliquities}.
This dichotomy has been attributed to changes in the tidal realignment timescales driven by the different stellar structures on either side of the Kraft break \citep[e.g.][]{Albrecht2012ObliquitiesMisalignments}.

Given \mystar's effective temperature and metallicity (see Table~\ref{tab:wasp-94_parameters}), the Kraft break should be around $\sim 6000$~K, placing \mystar in the hot, fast rotating regime.
Thus, \myplanet is consistent with the previously identified pattern of highly oblique planets being hosted by hot ($T_\text{eff} > 6000$~K) stars.

The full evolutionary history of the WASP-94 system required to produce the observed orbital configuration will be studied further in future work with the goal of placing additional constraints on the system parameters.

\subsection{Transmission spectrum}
Using observations of one transit with HARPS we find tentative evidence for Na in the atmosphere of \myplanet. Depending on the treatment of the Rossiter-McLaughlin correction, we find detection significances of 3.7--4.4$\sigma$ and 2.7--3.6$\sigma$. 

While we used state-of-the-art methods to correct for telluric and stellar effects, we still found unusual structures in the transmission spectrum instead of Gaussian-shaped Na doublet absorption lines. For this reason, we investigated whether varying planet and stellar parameters within their 1$\sigma$ uncertainties could account for the observed shape, e.g.\ velocities computed by our RM analysis, systemic velocity, etc. However, we did not find any evidence for any planet or stellar parameters to alter the shape of the transmission significantly enough to explain the W-shape. Therefore we conclude that the cloudy observing conditions and low flux in several frames are driving the noise structure in the transmission spectrum. 

\subsection{Atmospheric inferences combining with low resolution data}

We combined our high spectral resolution HARPS observations of the atmosphere of \myplanet\ with low resolution data from the literature \citep{Ahrer2022LRG-BEASTS:NTT/EFOSC2} to test whether we can achieve tighter constraints and a higher detection significance when running an atmospheric retrieval analysis on both simultaneously. 

We find that the abundance of Na and the cloud-top level did show tighter posterior distributions than each data set individually (see Fig.\,\ref{fig:retrieval-posteriors}).
However, the detection significance of Na performed by Bayesian model comparison against a retrieval without Na did not improve significantly. With our combined analysis we find a significance of 3.2$\sigma$, while the retrievals on the individual data sets result in 2.6$\sigma$ and 3.1$\sigma$ for NTT and HARPS, respectively. This only marginal increase in the detection of Na may be due to temperature degeneracies in the two data sets. While the low-resolution-only retrieval prefers a lower temperature, the high resolution one converges to a higher temperature. Consequently, the combined retrieval shows something in between, as shown in Table~\ref{tab:priors}, more akin to the equilibrium temperature of the planet, thereby not increasing the overall detection significance of Na given the slight tension between the data sets. That the temperature is degenerate with the relative optical depth of Na was also found by \citet{Pino2018Combining189733b} in their study combining low and high resolution observations of hot Jupiter HD\,189733b.

Nevertheless, these Na detection significances comply with the ones found using solely the HARPS spectrum and comparing the Na doublet to surrounding bands as discussed in the previous section (see also Section\,\ref{sec:sodium-depth}), validating our sodium detection. 

\subsection{Future Avenues}
Further transit observations are needed to fully characterise the shape and potential offset of the Na absorption feature in the atmosphere of this planet. \myplanet is an ESPRESSO GTO target thus we expect that the high-resolution signals from the planet's atmosphere will be further explored by this team in the near future. 

In addition, \myplanet is scheduled to be observed by JWST in June 2024 \citep[ID \#3154;][]{2023jwst.prop.3154A} to test C/O and metallicity predictions based on planet formation and migration models following its retrograde and misaligned orbit \citep[e.g.][]{Oberg2011TheAtmospheres,Madhusudhan2014TowardsMigration,Booth2017ChemicalDrift}. 

\section*{Acknowledgements}
E.A. acknowledges the financial support of the Department of Physics at the University of Warwick and the Max Planck Society. This work was supported by a UKRI Future Leaders Fellowship MR/T040866/1.
R.N. acknowledges support from UKRI/EPSRC through a Stephen Hawking Fellowship (EP/T017287/1).
C.H.M. gratefully acknowledges support from the Leverhulme Centre for Life in the Universe at the University of Cambridge.
P.J.W acknowledges support from the Science and Technology Facilities Council (STFC) under consolidated grants ST/T000406/1 and ST/X001121/1.
This work has made use of data from the European Space Agency (ESA) mission {\it Gaia} (\url{https://www.cosmos.esa.int/gaia}), processed by the {\it Gaia} Data Processing and Analysis Consortium (DPAC, \url{https://www.cosmos.esa.int/web/gaia/dpac/consortium}). This work has made use of the VALD database, operated at Uppsala University, the Institute of Astronomy RAS in Moscow, and the University of Vienna. 

\section*{Data Availability}
The raw and HARPS-pipeline processed data is available via the ESO archive, programme 097.C-1025(B), PI: Ehrenreich.
 



\bibliographystyle{mnras}
\bibliography{references.bbl} 


\bsp	
\label{lastpage}
\end{document}